\documentclass{PoS}
\title{Dyson's Instability in Lattice Gauge Theory }

\ShortTitle{Dyson's Instability in Lattice Theory }

\author{A. Bazavov $^b$, A. Denbleyker $^a$, Daping Du $^a$, \speaker{Y. Meurice } $^a$, A. Velytsky $^c$ and Haiyuan Zou $^a$\\
 
$^a$Department of Physics and Astronomy, The University of Iowa\\
Iowa City, Iowa 52242, USA\\
E-mail: \email{alan-denbleyker@uiowa.edu}\\
        E-mail: \email{daping-du@uiowa.edu}\\
        E-mail: \email{yannick-meurice@uiowa.edu}\\
        E-mail: \email{haiyuan-zou@uiowa.edu}\\
        \\
$^b$Physics Department, University of Arizona\\
Tucson, AZ 85721, USA\\
E-mail: \email{bazavov@physics.arizona.edu}\\
\\
$^c$Physics Department, Brookhaven National Laboratory\\Upton NY 11973, USA\\
E-mail: \email{vel@bnl.gov}
}

\abstract{
We discuss Dyson's argument that the vacuum is unstable under a 
change $g^2 \rightarrow - g^2$, in the context of lattice gauge theory. For compact
gauge groups, the partition function is well defined at negative $g^2$, but 
the average plaquette $P$ has a discontinuity when $g^2$ changes sign. 
This reflects a change of vacuum rather than a loss of vacuum. 
In addition, $P$ has poles in the complex $g^2$ plane, located at the complex zeros of the partition function (Fisher's zeros). We discuss the relevance of these singularities for lattice perturbation theory. We present new methods to locate 
Fisher's zeros using numerical values for the density of state in 
$SU(2)$ and $U(1)$ pure gauge theory. We briefly discuss similar issues for 
$O(N)$ nonlinear sigma models where the 
local integrals are also over compact spaces.}

\FullConference{The XXVII International Symposium on Lattice Field Theory - LAT2009\\
		 July 26-31 2009\\
		 Peking University, Beijing, China}

\begin{document}
\def\np{\mathcal{N}_p}
\section{Introduction}
Dyson instability \cite{dyson52,leguillou90} - the catastrophe happening when you 
change the sign of $e^2$ in QED - is often invoked to limit the validity of perturbation theory and justify the factorial growth of the perturbative coefficients. In the functional integral formulation of scalar models, this type 
of instability is related to large field 
configurations \cite {convpert,optim03}. 

For lattice models with compact field  integration (nonlinear sigma models over compact manifolds and lattice gauge theories (LGT) with compact groups), the large field problem is in principle absent. 
For $g^2<0$, the partition function is well defined and the change of sign of $g^2$ appears as a mere change in vacuum rather than a catastrophic instability. 
Can this explain the apparent power growth (rather than a factorial growth) observed in perturbative series for the average plaquette $P$ in Refs. \cite{direnzo2000,rakow2002,rakow05,Ilgenfritz:2009ck}? These series are consistent 
with the existence of Fisher's zeros (zeros of the partition function in the complex coupling plane) close to the real axis \cite{third}. It seems clear that a complete knowledge of the location of the Fisher's zero would provide a compete understanding of the complex singularities of $P$. The volume dependence of these zeros also provides important information regarding the order of possible transitions or the absence thereof
\cite{janke01,Ejiri:2005ts}. 

In these proceedings, we report recent results concerning these questions. 
Dyson instability is reviewed in Sec. \ref{sec:dyson} in the context of LGT. New ``topological" methods to locate Fisher's zeros in $SU(2)$ and $U(1)$ using numerical calculations of the density of states \cite{Denbleyker:2008ss,Bazavov:2009pj} are discussed in Sec. \ref{sec:fisher}. Similar questions for $O(N)$ sigma models in the complex t' Hooft coupling plane  are briefly discussed in Sec. \ref{sec:sigma}. 
Details can be found in a recent preprint \cite{Meurice:2009bq}. 

\section{Dyson's instability versus compact integration}
\label{sec:dyson}

Dyson's argument goes as follows \cite{dyson52,leguillou90}. 
Suppose that a physical quantity in QED can be calculated as a perturbative series  $F(e^2)= a_0+a_1e^2+\dots$ .
If we assume that the series has a finite radius of convergence, then, for $e^2$ sufficiently small,  we can interpret 
$F(-|e^2|)$ as the value of this quantity in a fictitious world where same charge particles attract. But in this fictitious world, every physical state is unstable. So, the radius of convergence is zero. 
Quoting the author  ``The argument [...] is lacking in mathematical rigor 
and in physical precision. It is intended to be suggestive, to serve as a basis for further discussions". 

The connection between asymptotic series and the problem of integrating large fields contributions can be understood with this very simple example
\begin{equation}
\int_{-\infty}^{+\infty}d\phi e^{-\frac{1}{2}\phi^2-\lambda \phi^4}\neq \sum_{q=0}^{\infty}
\frac{(-\lambda)^q}{q!} \int_{-\infty}^{+\infty}d\phi e^{-\frac{1}{2}\phi^2}\phi^{4q}\ .
\end{equation}
The sum and the integration have been interchanged illegally. 
The peak of the integrand of the $q$-th order term of the r.h.s is reached when $\phi^2=4q$. The approximation of $e^{-\lambda \phi^4}$ by an expansion of order $q$ in $\lambda\phi ^4$ is good provided that $\lambda\phi^4 <<q$, but at the peak of the integrand, $\phi^4=16q^2$ and we need $\lambda16q^2<<q$, which fails for $q$ large enough. 
On the other hand, if we introduce a field cutoff, as the order increases, at some order, the peak moves outside of the integration range and there is no factorial growth.
The general expectation is that  for a finite lattice, the partition function $Z$ calculated with a field cutoff is convergent and $\ln(Z)$ has a finite radius of convergence controlled by the zeros of the partition function. 
The field cutoff $\phi_{max}$ is an optimization parameter fixed using  strong coupling \cite{optim03}, for instance.

A fact that is obvious but which importance regarding weak coupling expansions may have been overlooked is  that 
lattice gauge theories with a {\it compact} group and nonlinear $O(N)$ sigma models have a {\it build-in large field cutoff}. In lattice gauge theory, the group elements associated with the links are integrated with $dU_l$ the compact Haar measure. 
Our notations are as follows: $N_c$ is the number of colors, 
$S=\sum_{plaq.}(1-(1/N_c)Re Tr(U_p))$ and 
$\beta=2N_c/g^2$. 
The number of plaquettes is denoted
$\mathcal{N}_p\equiv\ L^D D(D-1)/2\ $. The 
average plaquette: 
$P(\beta )\equiv(1/\mathcal {N}_p)\left\langle S\right\rangle $ will be our main object of study. 
\def\mn{\mathcal{N}_p}
The partition function $Z(\beta)$ is the Laplace transform of $n(S)$, the density of states: 
\begin{equation}
Z(\beta) =\int_0^{S_{max}}dS\ n(S)\ {\rm e}^{-\beta S}\ ,
\label{eq:intds}
 \end{equation}
 with
 \begin{equation}
n(S)=\prod_{l}\int dU_l \delta(S-\sum_{p}(1-(1/N_c)Re Tr(U_p)))\ .
\end{equation}
Assuming that ln($n(S)$) is extensive we can write 
\begin{equation}
\label{eq:colent}
n(S)={\rm e}^{\mn f(S/\mn)}
\ .
\end{equation} 
It is important to notice that at finite volume, $S_{max}$ is finite. For instance, 
$S_{max}=2\mn$ for $SU(2N)$ and $\frac{3}{2}\mn $ for $SU(3)$. 
In the strong coupling expansion, we expand in power of $\beta$: 
$Z=\sum_{n=0}^\infty z_n \beta^n$ with $|z_n|<S_{max}^n/n!$, so at finite volume, $Z$ is an analytical function, not only on the negative 
real axis, but over the entire $\beta$ plane. 

On the other hand, it is possible to show that for $SU(2N)$ on even lattices \cite{gluodyn04}
\begin{equation}
Z(-\beta)={\rm e}^{2\beta\mathcal{N}_p}Z(\beta)\  .
	\label{eq:su2id}
\end{equation}
Consequently, 
\begin{equation}
n(2\mn -S)=n(S)\  \  \  {\rm and }\   \   \  P(\beta)+P(-\beta)=2
\label{eq:dual}
\end{equation}
Since $lim_{\beta\rightarrow+\infty}P(\beta)=0$, $P$ has a discontinuity at $g^2=0$ and a regular series for $P$ about $g^2=0$ is not possible. However, it does not necessarily mean that 
the series has factorial growth. 

It is useful to consider first the case of a single  $SU(2)$ plaquette \cite{plaquette}. In that case, 
$n(S)=\frac{2}{\pi }\sqrt{S(2-S)}$   (invariant under $S\rightarrow2-S$). 
The large order of the weak coupling expansion $\beta\rightarrow +\infty$ is determined 
by the behavior of $n(S)$ near $S=2$, itself probed when $\beta\rightarrow -\infty$
in agreement with the common wisdom that the large order behavior 
of weak coupling series can be understood in terms of the behavior at small negative 
coupling.
$\sqrt{2-S}$ is then expended about $S=0$ (radius of convergence = 2). This yields the {\it convergent} expansion 
\begin{equation}
Z(\beta)=(\beta\pi)^{-3/2} 2^{1/2} 
	\sum_{l=0}^{\infty} (2\beta)^{-l}
\frac{\Gamma(l+1/2)}{l!(1/2-l)}\int_0^{2\beta}dt {\rm e}^{-t}t^{l+1/2}
\end{equation} 
As expected this is a not a regular series in the sense that the ``coefficients'' of $\beta^{-l}$ depend on $\beta$, but in a way that is invisible in perturbation theory. The crucial step is to get $\beta$-independent coefficients by neglecting the missing tails of integration. 
\begin{equation}
\int_0^{2\beta}dt {\rm e}^{-t}t^{l+1/2}\simeq \int_0^{\infty}dt {\rm e}^{-t}t^{l+1/2} + O({\rm e}^{-2\beta})
\end{equation}
 which in turn creates a factorial growth of the coefficients. 
The peak of the integrand crosses the boundary near order $2\beta$. 
Dropping higher order terms (than order $\simeq 2\beta$) agrees with the rule of thumb (minimizing the first contribution dropped). 
The non-perturbative part can be fully reconstructed (higher orders + tails) \cite{npp}. 
For  $L^4$ lattices, the crossing should be near order $2\beta\mn$. 
Non-perturbative effects should be explainable by the contributions near $S_{max}$. We plan to study this question on small lattices.
\section{Fisher's zeros from the density of states} 
\label{sec:fisher}
The poles of $P$ are located at  the Fisher's zeros. 
At finite volume, we expect these zeros to be isolated in the $\beta$ plane. It seems plausible that the zeros will accumulate along lines going through 0 in the $1/\beta$ plane as they do for Bessel functions. 
It is possible to use $n(S)$ to calculate $Z$ at complex $\beta$.  
The calculation of $n(S)$  for $SU(2)$ 
is discussed in Ref. \cite{Denbleyker:2008ss}. Additional checks were made by calculating the first three moments of $n(S)$. For $U(1)$ lattice gauge theory, multicanonical methods relying on the 
Biased Metropolis-Heatbath Algorithm \cite{BB2005} were used
 \cite{Bazavov:2009pj}. 
Using this $U(1)$ density of states, we have calculated  the plaquette 
distribution calculated at fixed $\beta$ and checked that there is an approximately symmetric double peak near $\beta=0.979$ for a $4^4$ lattice. 

\begin{figure}
\includegraphics[width=2.1in,angle=0]{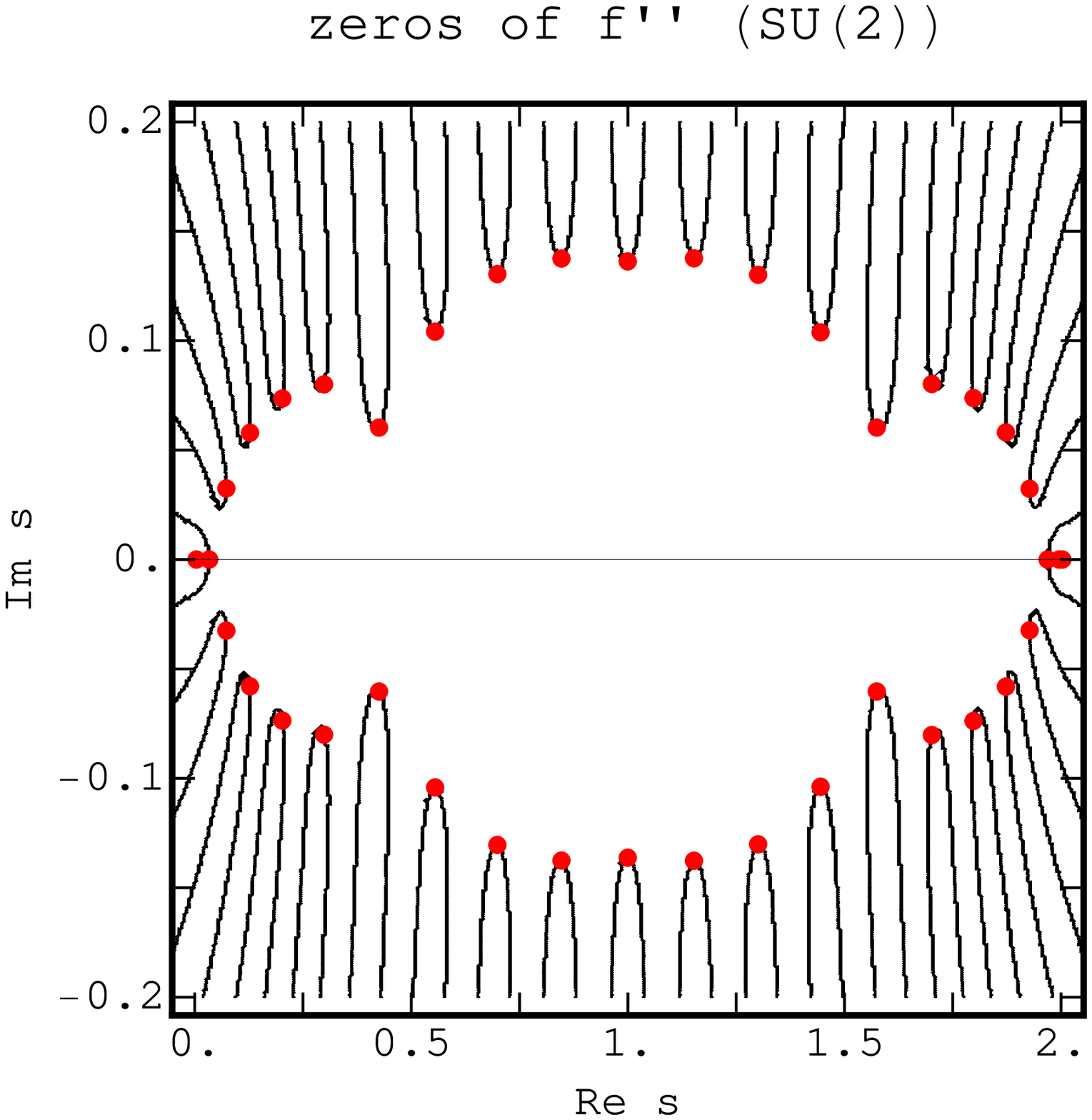}
\includegraphics[width=2.1in,angle=0]{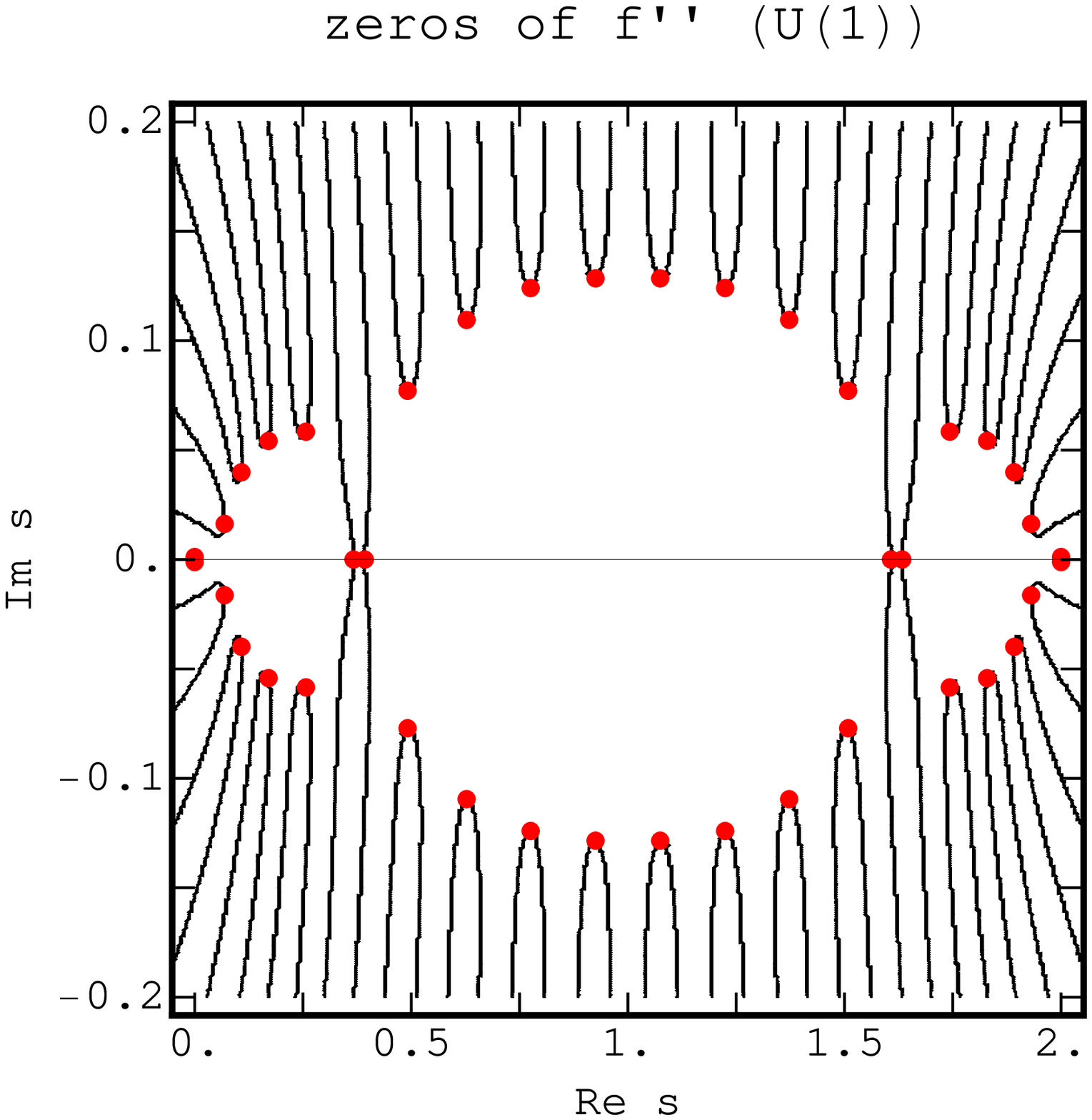}
\vskip30pt
\includegraphics[width=2.1in,angle=0]{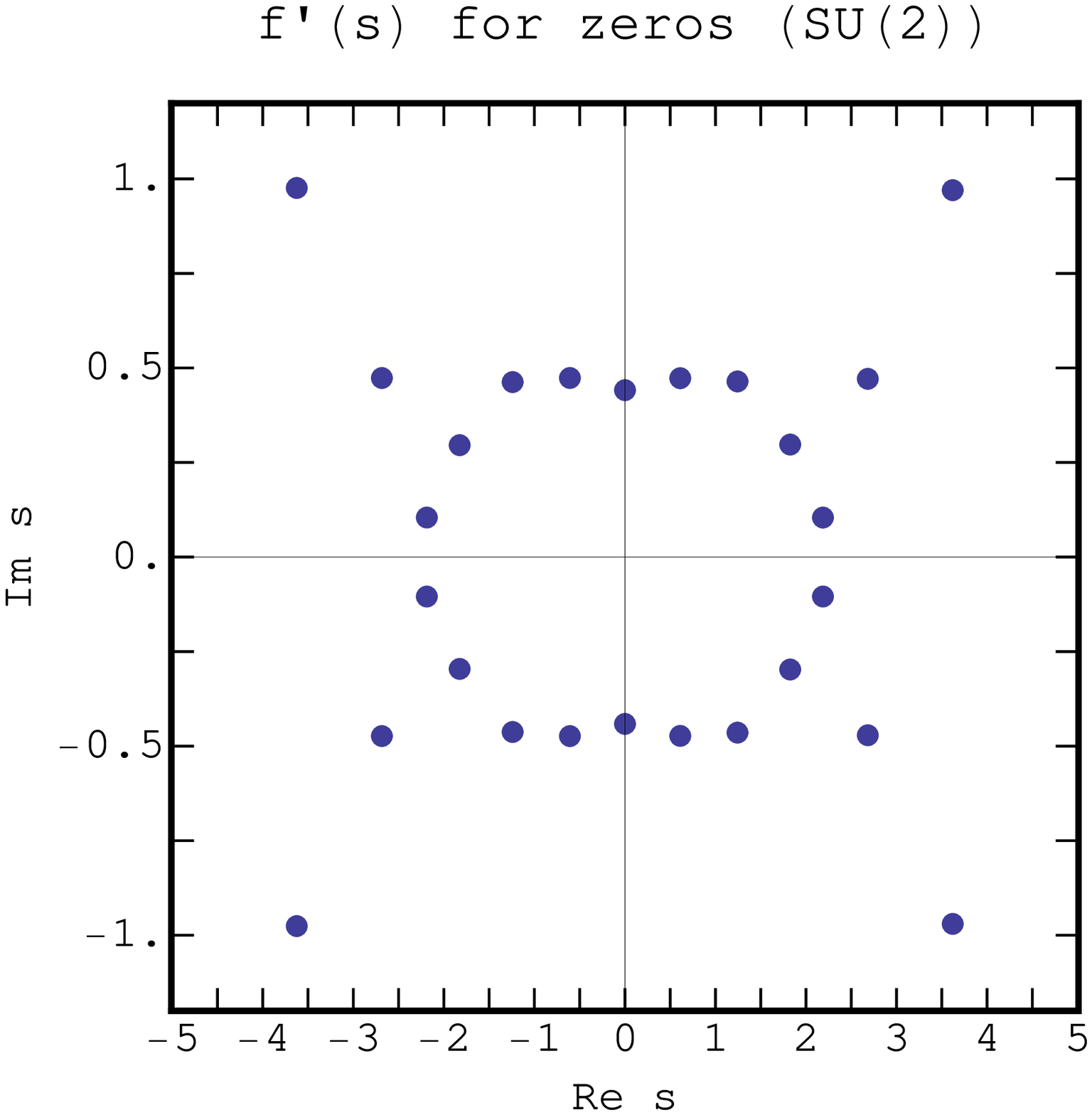}
\includegraphics[width=2.1in,angle=0]{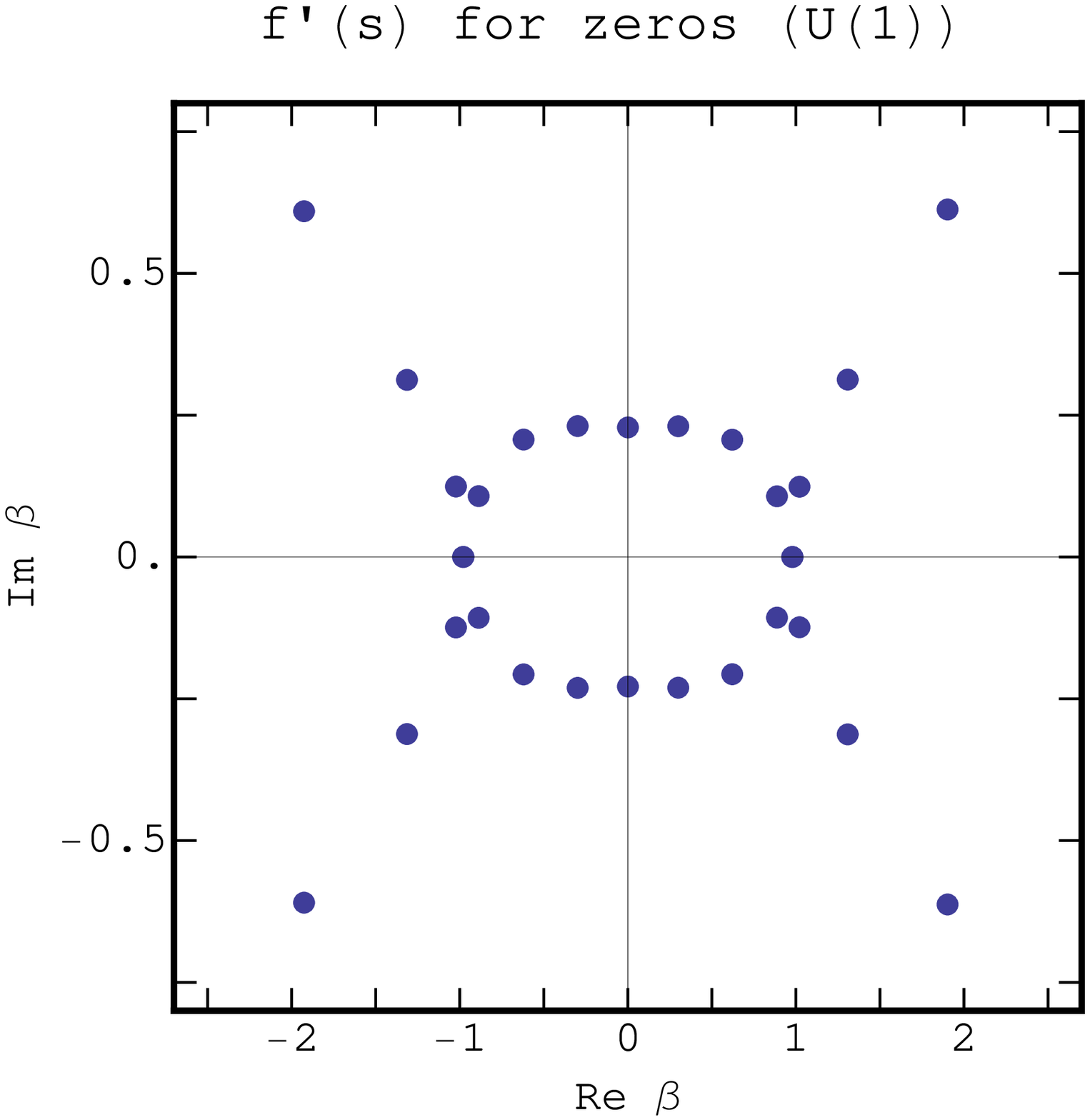}
\caption{\label{fig:fprime}Top:  complex zeros and zeros of the real part of $f''(s)$ in the complex $s$ plane with 40 Chebyshev polynomials on  $4^4$  for $SU(2)$ (left) and $U(1)$ (right).
Bottom: $f'(s)$ evaluated at the complex zeros of $f''(s)$ 
shown on the previous figure for $SU(2)$ (left) and $U(1)$ (right).}
\end{figure}

For both $U(1)$ and $SU(2)$ on a $4^4$ lattice, the numerical calculation of $Z(\beta)$ with $Im\beta\sim 0.2$ is difficult because $\beta$ is multiplied by $\mn$ and the 
integrand oscillate rapidly. A preliminary idea of the distribution of zero 
can be obtained using semi-classical methods. Using the ``color entropy" $f(s)$ defined in Eq. (\ref{eq:colent}), the saddle point of the 
integral is at $s_0$ given by solving $f'(s_0)=\beta$. Z becomes a Gaussian integral with correction of order $\sqrt{1/\mn}$ as long as $Re f''(s_0)<0$. As a Gaussian density of states has no complex zeros
\cite{alves91}, it seems clear that zeros should appear in regions of the $\beta$ plane corresponding to regions of the $s$ plane such that 
$Re f''(s_0)>0$. Using Chebyshev approximations of $f(s)$, we have 
constructed the boundary ($Re f''(s)=0$). The results are shown in Fig. \ref{fig:fprime}. The boundary form narrow tongues ending at a complex zeros of $f''$. These complex zeros are then mapped in the 
$\beta$ plane using $f'$. Their number depends on the degree of the 
polynomial approximation, but the general shape is robust under changes in the degree. It appears that in the case of $SU(2)$ the images in the $\beta$ plane are never on the real axis in contrast to the case of $U(1)$.

New methods have been developed to locate the Fisher zeros \cite{Meurice:2009bq,quasi}. Given the fact that $Z$ is an entire function in the $\beta$ plane, and that $P=-(dZ/d\beta)/Z$,  the worse thing that can happen to $P$ is that 
$Z$ has a zero of order $k$, say at $\beta_0$.
Then  $(dZ/d\beta)/Z\simeq k/(\beta-\beta_0)$ for $\beta\simeq \beta_0$. 
If we now integrate over a closed contour $C$, 
\begin{equation}
(i2\pi)^{-1}\oint_C  d\beta (dZ/d\beta)/Z = \sum_kn_k(C)\  ,
\label{eq:res}
\end{equation}
where $ n_k(C)$ is the number of zeros of order $k$ inside $C$
\def\nz{\sum_kn_k}. This allows us to monitor the accuracy of the calculation. We need to check that in good approximation, the real part is an integer and the imaginary part is zero. This is illustrated in  Fig. 
\ref{fig:jump} for a rectangular contour of variable height in the $\beta$ plane. Despite these encouraging results, there remain dependence on the 
interpolation or fit used to evaluate $f(s)$ numerically.   Resolving this issue should allow us to find finite size scaling for the zeros as discussed in Refs. \cite{janke01,Ejiri:2005ts}.
\begin{figure}
\includegraphics[width=2.8in,angle=0]{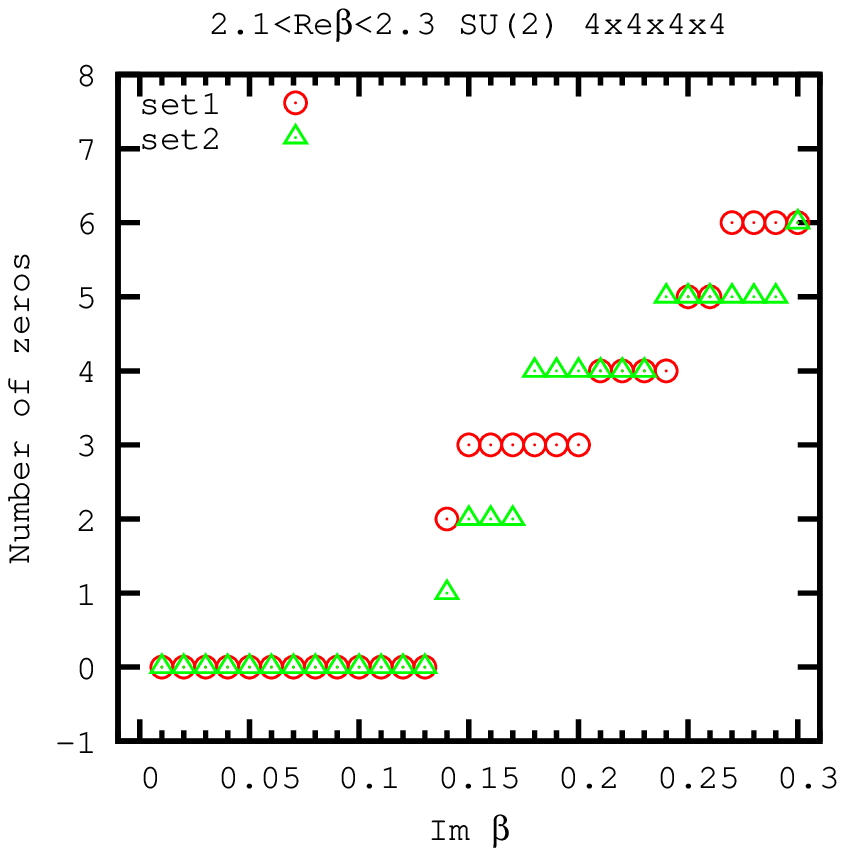}
\includegraphics[width=2.8in,angle=0]{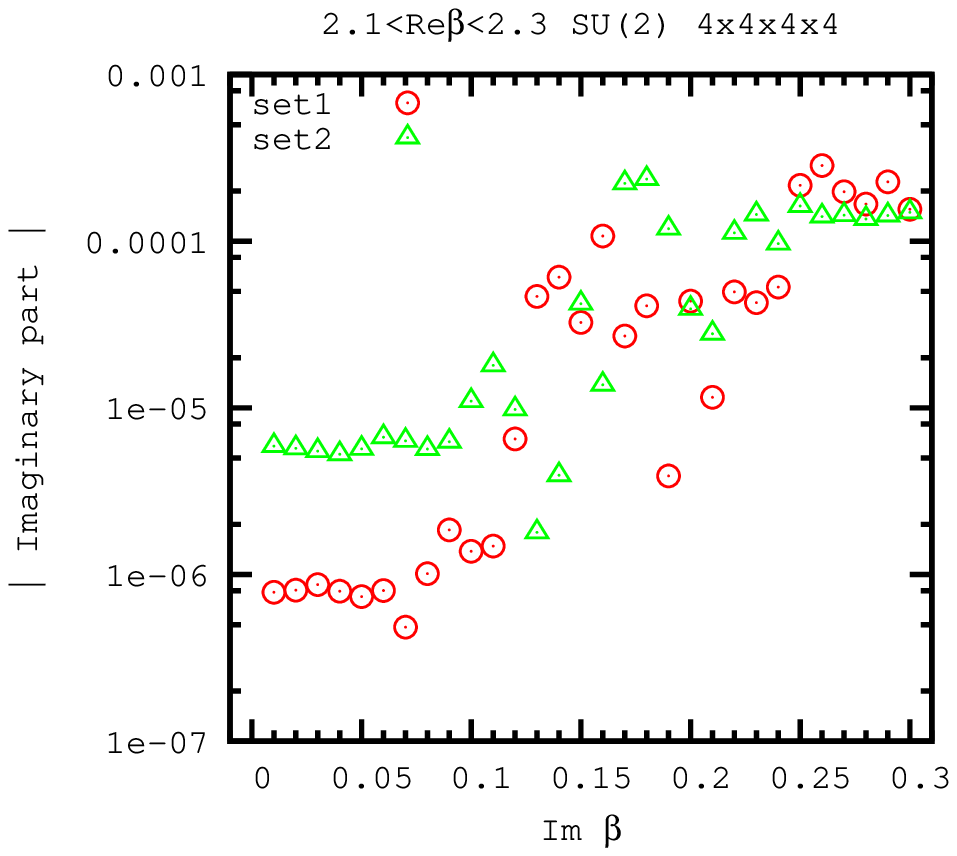}
\caption{ \label{fig:jump} 
$Re$ (left) and $Im$ (right) part of 
 $\nz $  defined in  Eq. (3.1) for a rectangular contour with 
$2.1<Re\beta<2.3$ and $0<Im\beta<y$ with a variable $y$, for 
$SU(2)$ on a $4^4$ lattice. Two independent numerical values of 
$n(S)$ were used. }
\end{figure}
\def\tc{\lambda^t}
\section{$2-D\ O(N)$ nonlinear sigma models}
\label{sec:sigma}
\begin{figure}
\includegraphics[width=2.4in,angle=0]{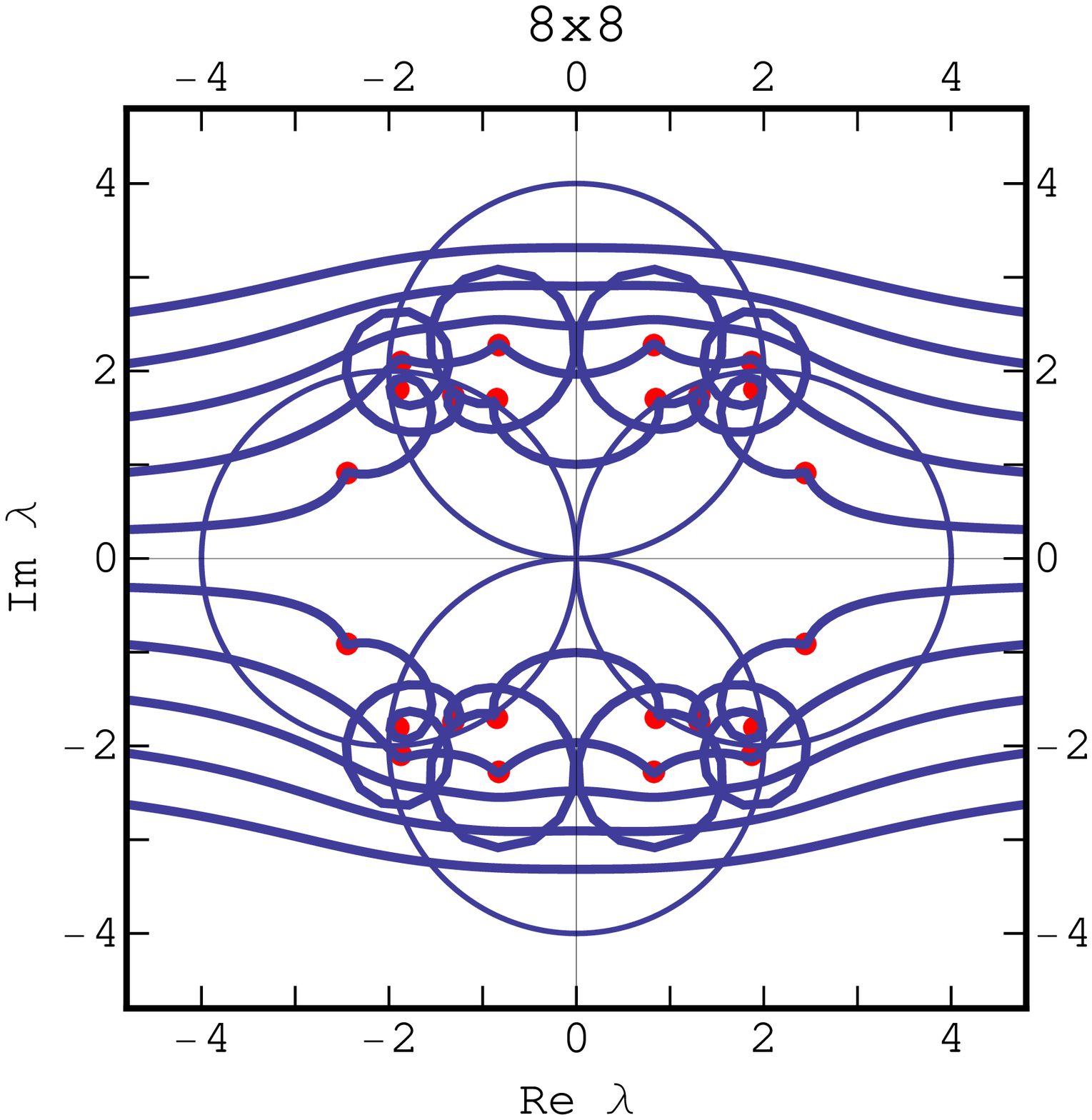}
\includegraphics[width=2.4in,angle=0]{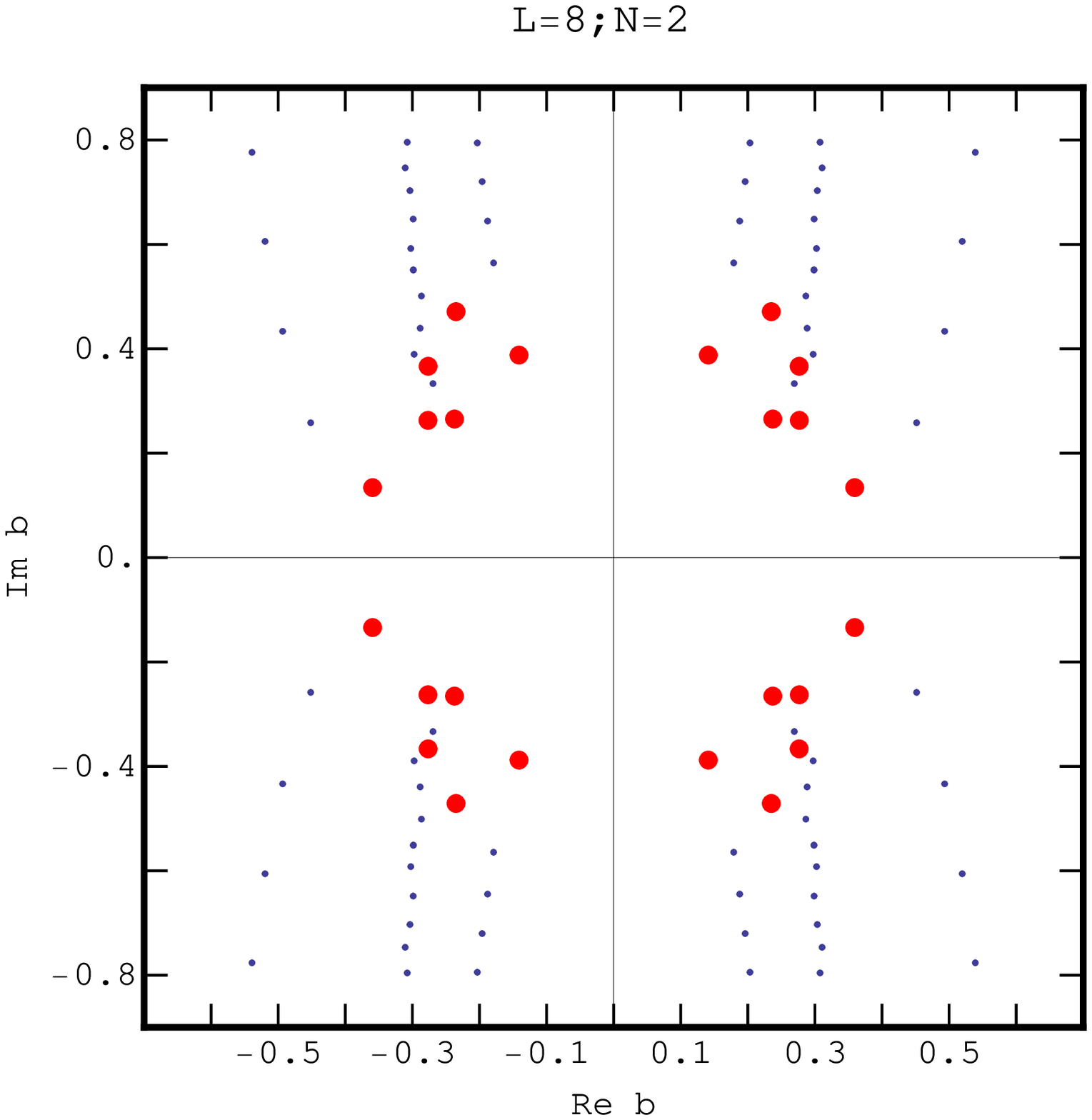}
\caption{\label{fig:fv}Left: images in the $\lambda ^t$ plane of lines of constant imaginary part 
 2.25, 1.75, 1.25, 0.75, 0.25, -0.25, ....,-2.25 in the complex mass gap plane and of the singular points (red dots) for a 
 8x8 lattice. Right: Fisher zeros for $N=2$ (blue) and images of singular points (red). }
\end{figure}

The nonlinear $O(N)$ sigma models on even cubic lattices have similar properties under the exchange of the sign of the coupling, namely $Z[-g^2]={\rm e}^{4DL^D/g^2}Z[g^2]$.  
The complex singularities of the average energy in 2 dimensions, for 
complex 't Hooft coupling $\tc= g_0^2N$ have been studied in the large-$N$ limit . 
Details can be found in a recent publication \cite{Meurice:2009bq}. 
A striking difference with the linear model is the absence of cut along the negative real axis. It was argued that  
the Fisher's zeros can only be inside 
a clover shaped region of the complex $\tc$ plane or equivalently outside of a region delimited by 4 
approximate hyperbolas with asymptotes on the boundary of a cross of width 0.5 centered at the origin in the  $1/\tc$ plane. The argument holds for large $N$ and large volume. This limit is being studied using 
exact results at finite $N$ and finite volume. The graphs of Fig. \ref{fig:fv} made with $N=2$ on 
a $8^2$ lattice can be compared with the corresponding ones in Ref. 
\cite{Meurice:2009bq}.
\section{Conclusions}
 For pure gauge models with compact groups, there is no loss of vacuum when $g^2_0\rightarrow -g^2_0$, but only a change of vacuum.
The discontinuity of the plaquette forbids the existence of a converging 
perturbative series but does not dictate the large order behavior. 
Reliable methods to locate Fisher's zeros are in progress. 
Non-perturbative effects should be accountable by modified expansions. 
New data for perturbative coefficients should help in this task . 
\begin{acknowledgments}
This 
research was supported in part  by the Department of Energy
under Contract No. FG02-91ER40664. Y. Meurice thanks H. Perlt for 
valuables discussions on new data. 
\end{acknowledgments}


\begin{thebibliography}{10}

\bibitem{dyson52}
F.~Dyson, Divergence of Perturbation Theory in Quantum Electrodynamics, 
\newblock {\em Phys. Rev.} {\bf 85}  631 (1952).

\bibitem{leguillou90}
J.~C. LeGuillou and J.~Zinn-Justin, 
\newblock {\em Large-Order Behavior of Perturbation Theory}.
\newblock North Holland, Amsterdam, 1990.

\bibitem{convpert}
Y.~Meurice, 
\newblock A simple method to make asymptotic series of Feynman diagrams
  converge,
\newblock {\em Phys. Rev. Lett.}  {\bf 88} 141601 (2002).

\bibitem{optim03}
B.~Kessler, L.~Li, and Y.~Meurice,
\newblock New optimization methods for converging perturbative series with a
  field cutoff,
\newblock {\em Phys. Rev.} {\bf D69} 045014 (2004).

\bibitem{direnzo2000}
F.~Di~Renzo and L.~Scorzato,
\newblock A consistency check for renormalons in lattice gauge theory:
  $\beta^{-10}$ contributions to the su(3) plaquette,
\newblock {\em JHEP} {\bf 10} 038 (2001).

\bibitem{rakow2002}
R.~Horsley, P.~E.~L. Rakow, and G.~Schierholz,
\newblock Separating perturbative and non-perturbative contributions to the
  plaquette.
\newblock {\em Nucl. Phys. Proc. Suppl.} {\bf 106} 870 (2002).

\bibitem{rakow05}
Paul E.~L. Rakow, 
\newblock Stochastic perturbation theory and the gluon condensate, 
\pos{PoS{LAT2005} 284} (2006).

\bibitem{Ilgenfritz:2009ck}
E.~M. Ilgenfritz et~al.,
\newblock {Wilson loops in very high order lattice perturbation theory},
arXiv:0910.2795, poster presented by H. Perlt at this conference .

\bibitem{third}
L.~Li and Y.~Meurice, 
\newblock About a possible 3rd order phase transition at t = 0 in 4d
  gluodynamics, 
\newblock {\em Phys. Rev.} {\bf D73} 036006 (2006).

\bibitem{janke01}
Wolfhard Janke and Ralph Kenna, 
\newblock Phase transition strengths from the density of partition function
  zeroes, 
\newblock {\em Nucl. Phys. Proc. Suppl.} {\bf 106} 905  (2002).

\bibitem{Ejiri:2005ts}
Shinji Ejiri, 
\newblock {Lee-Yang zero analysis for the study of QCD phase structure}, 
\newblock {\em Phys. Rev.} {\bf D73} 054502  (2006).

\bibitem{Denbleyker:2008ss}
A.~Denbleyker, Daping Du, Yuzhi Liu, Y.~Meurice, and A.~Velytsky, 
\newblock {Series expansions of the density of states in SU(2) lattice gauge
  theory}, 
\newblock {\em Phys. Rev.} {\bf D78} 054503 (2008).

\bibitem{BB2005}
  A.~Bazavov and B.~A.~Berg,
  Heat bath efficiency with Metropolis-type updating, 
  Phys.\ Rev.\  D {\bf 71} 114506 (2005)


\bibitem{Bazavov:2009pj}
Alexei Bazavov and Bernd~A. Berg, 
\newblock {Program package for multicanonical simulations of U(1) lattice gauge
  theory}, 
\newblock {\em Comput. Phys. Commun.} {\bf 180} 2339 (2009).

\bibitem{Meurice:2009bq}
Y.~Meurice, 
\newblock {Dyson instability for 2D nonlinear O(N) sigma models}, 
\newblock {\em Phys. Rev.} {\bf D80} 054020 (2009).

\bibitem{gluodyn04}
L.~Li and Y.~Meurice, 
\newblock Lattice gluodynamics at negative $g^2$, 
\newblock {\em Phys. Rev.} {\bf D71} 016008 (2005).

\bibitem{plaquette}
L.~Li and Y.~Meurice, 
\newblock An example of optimal field cut in lattice gauge perturbation theory, 
\newblock {\em Phys. Rev.} {\bf D71} 054509 (2005).

\bibitem{npp}
Y.~Meurice, 
\newblock {The non-perturbative part of the plaquette in quenched QCD},
\newblock {\em Phys. Rev.} {\bf D74} 096005 (2006).

\bibitem{alves91}
Nelson~A. Alves, Bernd~A. Berg, and Sergiu Sanielevici, 
\newblock Spectral density study of the su(3) deconfining phase transition, 
\newblock {\em Nucl. Phys.} {\bf B376} 218 (1992).


\bibitem{quasi}
A. Denbleyker, D. Du, and Y. Meurice, and A.~Velytsky, {\it Fisher's zeros of quasi-Gaussian densities of states}, Phys. Rev. 
{\bf D76} 116002 (2007).



\end{thebibliography}
\end{document}